\documentclass[11pt]{article}
\usepackage{amssymb,amsmath,psfrag}

\def \beq  {\begin{equation}}
\def \eeq  {\end{equation}}
\def \ba  {\begin{eqnarray}}
\def \ea  {\end{eqnarray}}
\def \non {\nonumber}

\def \CF {{\cal F}}
\def \CG {{\cal G}}
\begin{document}

\thispagestyle{empty}

\rightline{IPhT-T10/046}

\begin{center}
\vskip 2 cm
{\Large \bf On the two-loop hexagon Wilson loop remainder function\\
\vskip 3mm
in $N=4$ SYM}\\

\vskip 1.2 cm
{Jian-Hui Zhang}\\

\vskip 4mm
{\it Institut de Physique Th\'eorique, \\
CEA Saclay, \\
91191 Gif-sur-Yvette Cedex, France} \\
\vskip 1.5 cm

\end{center}

\begin{abstract}
A duality relation has been proposed between the planar gluon MHV amplitudes and light-like Wilson loops in $N=4$ super Yang-Mills. At six-point two-loop, the results for the planar gluon MHV amplitude and for the light-like Wilson loop agree, but they both differ from the Bern-Dixon-Smirnov ansatz by a finite remainder function. Recently Del Duca, Duhr and Smirnov presented an analytical result for the two-loop hexagon Wilson loop remainder function in general kinematics. Their result is rather lengthy, and the dependence on the conformal cross ratios appears in a complicated way. Here we present an alternate, more compact representation for the two-loop hexagon Wilson loop remainder function.

\end{abstract}

\newpage
\setcounter{page}{1}\setcounter{footnote}{0}

\section{Introduction}
In the past few years much progress has been made in understanding scattering amplitudes in gauge theories, and in particular in $N=4$ supersymmetric Yang-Mills (SYM) theory. An interesting feature of the $N=4$ SYM amplitudes is the all-loop iterative structure proposed by Bern, Dixon and Smirnov (BDS)~\cite{Bern:2005iz} for the maximally-helicity-violating (MHV) planar gluon amplitudes. Their proposal is based on the observation of an iteration relation between one- and two-loop planar four-gluon MHV amplitudes by Anastasiou, Bern, Dixon and Kosower (ABDK)~\cite{Anastasiou:2003kj} and an explicit computation of the four-gluon amplitude at three loops in ref.~\cite{Bern:2005iz}.

The planar gluon amplitudes can be factorized into a universal infrared (IR) divergent factor and a finite part. Given the well-known structure of the IR divergences of gluon amplitudes~\cite{IRrefs}, the BDS ansatz proposes an explicit expression for the finite part of the planar gluon MHV amplitude with an arbitrary number of external gluons, to all orders in 't Hooft coupling. An important aspect of the ansatz is that the kinematic dependence of the finite part is described by a function whose coupling dependence can be factored out, and the remaining coupling-independent part of the function is given by the finite part of the box functions entering one-loop MHV amplitude. Besides the tests for four-gluon amplitude up to three loops in refs.~\cite{Bern:2005iz,Anastasiou:2003kj}, the BDS ansatz has been shown to be correct also for two-loop five-gluon amplitude~\cite{Cachazo:2006tj,Bern:2006vw}.

In a remarkable paper~\cite{Alday:2007hr}, Alday and Maldacena were able to compute the planar gluon amplitudes at strong coupling using the AdS/CFT correspondence. Their result agrees with the strong coupling limit of the BDS ansatz for the four-gluon case. However, for amplitudes with a large number of external gluons, a discrepancy was found between the strong coupling prediction and the BDS ansatz~\cite{Alday:2007he}. This indicates a potential failure of the BDS ansatz for amplitudes with a sufficiently large number of external gluons.

Alday and Maldacena also pointed out that in the strong coupling limit the computation of planar gluon amplitudes is equivalent to the computation of the vacuum expectation value of polygonal Wilson loops with light-like edges defined by the momenta of external gluons. This suggests a duality between planar gluon amplitudes and light-like Wilson loops at strong coupling. Such a duality was then conjectured to hold at weak coupling~\cite{Drummond:2007aua}, and was verified by explicit one-loop computations for four-sided Wilson loop in the same paper and for an arbitrary n-sided case in ref.~\cite{Brandhuber:2007yx}. Further two-loop results for four-, five- and six-sided Wilson loops~\cite{Drummond:2007cf,Drummond:2007au,Drummond:2007bm,Drummond:2008aq} also found agreement with the gluon amplitude results~\cite{Bern:2005iz,Anastasiou:2003kj,Cachazo:2006tj,Bern:2006vw,Bern:2008ap}. At six-point two-loop level, both the Wilson loop and the amplitude results differ from the BDS ansatz.

The light-like $n$-sided Wilson loop exhibits an anomalous conformal symmetry, and the associated anomalous conformal Ward identities constrain the form of the light-like Wilson loop~\cite{Drummond:2007au}. In general, the solution of the anomalous conformal Ward identities is uniquely determined up to a function invariant under the conformal symmetry. Such a function can be constructed from the conformally-invariant cross ratios built out of the external momenta. For $n\le5$, it is not possible to construct such conformal cross ratios due to the light-likeness of the external momenta. Given the duality between planar gluon amplitudes and light-like Wilson loops (which implies that the planar gluon amplitude also satisfies the anomalous conformal Ward identities), this provides an explanation for the correctness of the BDS ansatz for four- and five-gluon amplitudes: it satisfies the anomalous conformal Ward identities, and is unique because of the lack of conformal cross ratios. For $n\ge6$, the conformal cross ratios can be constructed (for $n=6$, there are $3$ such ratios), therefore a solution to the anomalous conformal Ward identities can differ from the BDS form by a function of the conformal cross ratios.

Explicit numerical computations for the two-loop hexagon Wilson loop~\cite{Drummond:2008aq,Anastasiou:2009kna} showed that the complete result differs from the BDS ansatz by a finite remainder function, which depends only on the conformal cross ratios. In ref.~\cite{Anastasiou:2009kna}, also the two-loop seven- and eight-sided Wilson loops were evaluated numerically, and the corresponding remainder functions were shown to depend on the conformal cross ratios only.

In addition to the numerical computation, analytical evaluations of the remainder function have also been carried out recently, both at strong and weak couplings. In ref.~\cite{Alday:2009yn}, the remainder function was evaluated analytically at strong coupling in a special kinematic regime where only even-sided Wilson loops are admitted and where the number of independent cross ratios is reduced so that the function is non-trivial only for $n\ge 8$. The explicit form of the octagon Wilson loop remainder function in this special kinematics was also given there. Later on, a numerical evaluation of the octagon Wilson loop remainder function at two loops was carried out~\cite{Brandhuber:2009da} in the same kinematics and compared to the strong coupling  result. The numerical comparison suggests a linear relation between the remainder functions at weak and strong couplings. In ref.~\cite{Alday:2009dv}, Alday, Gaiotto and Maldacena computed analytically the hexagon Wilson loop remainder function in a kinematic regime where the three conformal cross ratios are equal, and they found a fairly simple functional form. Recently, Del Duca, Duhr and Smirnov presented an analytical result for the two-loop hexagon Wilson loop remainder function at weak coupling~\cite{DelDuca:2009au,DelDuca:2010zg}, starting from the Feynman integrals contributing to the two-loop hexagon Wilson loop given in ref.~\cite{Anastasiou:2009kna}. They considered the hexagon Wilson loop in the quasi-multi-Regge kinematics~\cite{Fadin:1989kf,DelDuca:1995ki} where the Wilson loop exhibits exact Regge factorization\footnote{An exact Regge factorization of four-sided Wilson loop has been observed in ref.~\cite{Drummond:2007aua}.}, therefore the analytic dependence of the remainder function on the conformal cross ratios is not modified by going to this kinematics, but the computation of the remainder function simplifies remarkably.

The two-loop hexagon Wilson loop remainder function was expressed in refs.~\cite{DelDuca:2009au,DelDuca:2010zg} in terms of transcendental weight four terms constructed from Goncharov polylogarithms and harmonic polylogarithms. The result is rather lengthy and the dependence on the conformal cross ratios appears in a complicated way\footnote{By evaluating the integrals contributing to the two-loop hexagon Wilson loop collected in ref.~\cite{Drummond:2008aq}, one ends up with an expression~\cite{Korchemsky:2010unpub} for the remainder function of comparable size to that in refs.~\cite{DelDuca:2009au,DelDuca:2010zg}.}. In order to extract interesting physical information from the remainder function and to eventually find a systematic way to fix the BDS ansatz, a fairly simple representation of the remainder function is desirable. Moreover, the numerical results of ref.~\cite{Brandhuber:2009da} suggest a potential link between the remainder functions at weak and strong couplings. Given the simplicity of the hexagon Wilson loop remainder function at strong coupling~\cite{Alday:2009dv}, one desires to have a simple representation also for the hexagon Wilson loop remainder function at weak coupling, in order to make a comparison with the strong coupling result. In this paper we present an alternate, more compact representation for the two-loop hexagon Wilson loop remainder function, based on the observation that the conformal-cross-ratio-dependent terms in the BDS ansatz exhibit a simple structure when written in an integral form, and on the result of~\cite{DelDuca:2009au} as well as on the general properties of multiple polylogarithms described in refs.~\cite{Gehrmann:2000zt,Vollinga:2004sn}.

The paper is organized as follows. In Section~\ref{duality} we briefly review the proposed duality between the planar gluon MHV amplitude and the light-like Wilson loop, and the definition of remainder function. In Section~\ref{remfunc62} we present our representation for the two-loop hexagon Wilson loop remainder function, both in the kinematic configurations where the three conformal cross ratios coincide and in general kinematics. We also briefly discuss the analytic properties of the remainder function. Our conclusion is given in Section~\ref{conclusion}.

\section{The planar gluon amplitude/Wilson loop duality and the remainder function}\label{duality}
The color-ordered planar gluon MHV amplitude in $N=4$ SYM can be factorized and written as
\beq\label{logamp}
\ln\mathcal{M}_n^{\rm (MHV)}=Z_n^\text{IR, div}+F_n^{\rm (MHV)}(p_1,\ldots,p_n;a)+\mathcal O(\epsilon)\ ,
\eeq
where the left-hand side is the logarithm of the rescaled amplitude, defined as the ratio of the color-ordered amplitude and the corresponding tree amplitude. $Z_n^\text{IR, div}$ represents the IR divergences of the amplitude, regularized by dimensional reduction in $D=4-2\epsilon$ dimensions. $F_n^{\rm (MHV)}$ is the finite part depending on the momenta of external gluons $p_i (i=1,\ldots,n)$ and on the 't Hooft coupling $a=g^2 N/(8\pi^2)$. In $N=4$ SYM, the IR divergence is characterized by the universal cusp anomalous dimension (for the leading poles) and the collinear anomalous dimension (for the subleading poles). The finite part $F_n^{\rm (MHV)}$ can be extracted from direct computation of planar $n$-gluon MHV amplitude, which can be carried out, e.g. with the unitarity-cut techniques~\cite{Bern:1994zx,Bern:1994cg}; and compared with the BDS proposed form.

On the other hand, the light-like Wilson loop dual to the planar gluon amplitude is given by
\beq
W(C_n)=\frac{1}{N}\left<0|\right.\,{\rm Tr}\, {\rm P} \exp\left(i\oint_{C_n} dx^\mu A_\mu(x)\right)\left.|0\right>\ ,
\eeq
where the gauge fields $A_\mu(x)$ are integrated along the light-like polygonal contour $C_n$ with $n$ cusps $x_i^\mu$, the difference of which is given by the momenta of external gluons in the dual planar gluon amplitude as
\beq
x_i^\mu-x_{i+1}^\mu=x_{i,i+1}^\mu=p_i^\mu\ .
\eeq
The Wilson loop defined above has ultraviolet (UV) or cusp divergences, due to the presence of cusps on the integration contour $C_n$.

The Wilson loop can also be factorized into a UV divergent part and a finite part~\cite{Drummond:2008aq}. Its logarithm can be written as
\beq\label{logWL}
\ln W(C_n)=Z_n^\text{UV, div}+F_n^{\rm (WL)}(x_1,\ldots,x_n;a)+\mathcal O(\epsilon)\ .
\eeq
The leading UV divergence of the Wilson loop (regularized by dimensional reduction) is characterized by the cusp anomalous dimension. The fact that it also appears in the IR divergent part of the amplitude reflects the relation between IR divergences of scattering amplitudes and UV divergences of Wilson loops with cusps~\cite{IRUV}. It has been shown~\cite{Drummond:2008aq} that, with an appropriate identification of the respective regularization parameters, the IR divergences of planar gluon amplitudes and the UV divergences of Wilson loops match with each other, the proposed planar gluon amplitude/Wilson loop duality then amounts to an equality of the finite parts in Eqs.~(\ref{logamp}) and (\ref{logWL}) (up to an irrelevant constant)
\beq\label{dualityfinpart}
F_n^{\rm (MHV)}=F_n^{\rm (WL)}+\text{const.}
\eeq

To verify this duality relation, one needs to perform the calculation of light-like Wilson loops. This can be significantly simplified by making use of the non-abelian exponentiation properties of Wilson loops~\cite{Gatheral:1983cz,Frenkel:1984pz}, which allow one to write the Wilson loop as an exponential, and its logarithm can be written as
\beq\label{WLexponential}
\ln W(C_n)=\ln\Big(1+\sum_{l=1}^\infty a^l W_n^{(l)}\Big)=\sum_{l=1}^\infty a^l w_n^{(l)}\ .
\eeq
Expanding Eq.~(\ref{WLexponential}) to two-loop order, one finds
\beq
w_n^{\rm (1)}=W_n^{\rm {1}}\ ,\hspace{5em} w_n^{\rm (2)}=W_n^{\rm (2)}-\frac{1}{2}(W_n^{\rm (1)})^2\ .
\eeq
The one- and two-loop Wilson loop coefficients $w_n^{\rm (1)}, w_n^{\rm (2)}$ have been computed in~\cite{Drummond:2007aua,Brandhuber:2007yx,Drummond:2007cf,Drummond:2007au,Drummond:2008aq,Anastasiou:2009kna}, where for $n\ge6$ at two-loop only numerical results are available.

As mentioned before, the BDS ansatz proposes an explicit expression for the finite part of the planar gluon MHV amplitude, $F_n^{\rm MHV}$. For the six-gluon amplitude, the ansatz gives
\begin{align}
F_6^{\rm (BDS)}&=\frac{1}{4} \Gamma_{\rm cusp}(a)\sum_{i=1}^6\Big[-\ln\Big(\frac{x_{i,i+2}^2}{x_{i,i+3}^2}\Big)\ln\Big(\frac{x_{i+1,i+3}^2}{x_{i,i+3}^2}\Big)+\frac{1}{4}\ln^2\Big(\frac{x_{i,i+3}^2}{x_{i+1,i+4}^2}\Big)-\frac{1}{2}{\rm Li_2}\Big(1-\frac{x_{i,i+2}^2 x_{i+3,i+5}^2}{x_{i,i+3}^2 x_{i+2,i+5}^2}\Big)\Big]\non\\
&+\text{const.}\non
\end{align}
\begin{align}\label{6gBDS}
&=\frac{1}{4} \Gamma_{\rm cusp}(a)\Big\{\sum_{i=1}^6\Big[-\ln\Big(\frac{x_{i,i+2}^2}{x_{i,i+3}^2}\Big)\ln\Big(\frac{x_{i+1,i+3}^2}{x_{i,i+3}^2}\Big)+\frac{1}{4}\ln^2\Big(\frac{x_{i,i+3}^2}{x_{i+1,i+4}^2}\Big)\Big]-\sum_{i=1}^3{\rm Li_2}(1-u_i)\Big\}+\text{const.}
\end{align}
where $\Gamma_{\rm cusp}(a)$ is the cusp anomalous dimension with the perturbative expansion $\Gamma_{\rm cusp}(a)=2a-2\zeta_2 a^2+\mathcal O(a^3)$. In the second equation above we have rewritten the last term in the bracket in terms of the three conformal cross ratios defined as
\beq
u_1=\frac{x_{13}^2 x_{46}^2}{x_{14}^2 x_{36}^2}\ ,\hspace{3em} u_2=\frac{x_{24}^2 x_{15}^2}{x_{25}^2 x_{14}^2}\ ,\hspace{3em} u_3=\frac{x_{35}^2 x_{26}^2}{x_{36}^2 x_{25}^2}\ .
\eeq
It is the simplicity of the conformal-cross-ratio-dependent term ${\rm Li_2}(1-u_i)$ that guides us in seeking a simple representation of the remainder function.

The amplitude remainder function is defined as the difference between the logarithm of the rescaled amplitude and the corresponding BDS ansatz, and it has a trivial behavior under collinear limits~\cite{Bern:2008ap}. Given the duality relation Eq.~(\ref{dualityfinpart}), the Wilson loop remainder function can be defined as~\cite{Drummond:2008aq}
\beq
R_n^{\rm WL}=F_n^{\rm (WL)}-F_n^{\rm (BDS)}\ .
\eeq
The two-loop hexagon Wilson loop remainder function $R_6^{\rm WL, (2)}$ is then obtained by computing the two-loop contribution to $F_6^{\rm (WL)}$. This definition for $R_6^{\rm WL, (2)}$ differs by a constant from the one used in ref.~\cite{Anastasiou:2009kna}, where the Wilson loop remainder function is defined such that it has precisely the same collinear behavior as the amplitude remainder function. Here we follow the definition for $R_6^{\rm WL, (2)}$ of~\cite{Anastasiou:2009kna}, as was used in refs.~\cite{DelDuca:2009au,DelDuca:2010zg}. Note that $R_6^{\rm WL, (2)}$ is a function of $u_1, u_2, u_3$ only.

\section{The two-loop hexagon Wilson loop remainder function}\label{remfunc62}
In ref.~\cite{DelDuca:2009au}, an analytical result for the two-loop hexagon Wilson loop remainder function was presented, which was expressed in terms of transcendental weight four terms constructed from Goncharov polylogarithms and harmonic polylogarithms, in accord with the expectation based on transcendentality arguments that perturbative corrections to the Wilson loop at $L$-loop should have transcendentality $2L$~\cite{Drummond:2007bm}. The result of~\cite{DelDuca:2009au} is rather lengthy and the dependence of the remainder function on the conformal cross ratios appears in a complicated way.
In order to extract interesting physical information from the remainder function and to eventually fix the BDS ansatz, and also to shed light on the potential connection between the remainder functions at weak and strong couplings, a fairly simple representation of the remainder function is strongly desirable.

Although at two-loop the BDS ansatz for the six-gluon MHV amplitude fails to  produce the complete result, its form provides some hint for seeking a simple representation of the remainder function. As one can see from Eq.~(\ref{6gBDS}), the two-loop contribution to $F_6^{\rm (BDS)}$ contains terms that are functions of the three conformal cross ratios only, i.e. the dilogarithms ${\rm Li_2}(1-u_i)$ (they have transcendentality two, the remaining transcendentality is provided by $\Gamma_{\rm cusp}$). Such terms have the following simple integral form
\beq
\zeta_2\,{\rm Li_2}(1-u_i)=-\int_0^1\frac{\ln(1-(1-u_i)t)\,\zeta_2}{t}\ ,
\eeq
where $\zeta_2=\pi^2/6$ is the Riemann zeta constant. Inspired by this, we try to find a simple representation for the two-loop hexagon Wilson loop remainder function, which comprises terms that can all be written as the following form
\beq\label{intrep}
\int_0^1\frac{\ln(a-b\,t)R_2(t)}{t-c}\ ,
\eeq
where $R_2$ is a transcendentality two function of the three conformal cross ratios $u_i$ (as well as of the integration parameter). Namely, it contains, in addition to the transcendentality two constant $\pi^2$, only double logarithms and dilogarithms. The quantities $a, b, c$ depend on $u_i$ and have transcendentality zero. As we will see below, most terms of the function $R_2$ are double logarithms, only a small number of dilogarithms show up in $R_2$. In addition, in contrast to the result in ref.~\cite{DelDuca:2009au}, where the dependence through the square-root-containing variables on $u_i$ (for the definition of such variables see Eq.~(4.3) of ref.~\cite{DelDuca:2009au}) is very complicated, such variables only show up in the simple logarithm in Eq.~(\ref{intrep}) in the present representation. Therefore it is possible to combine them so that the square roots drop out.

In the following we present the representation of $R_6^{\rm WL, (2)}$ in the form of Eq.~(\ref{intrep}), both for the special case where all three conformal cross ratios are equal and for the general case. The derivation is based on the discussions above, and on the result of ref.~\cite{DelDuca:2009au} as well as on the general properties of multiple polylogarithms described in refs.~\cite{Gehrmann:2000zt,Vollinga:2004sn}.

We begin with the special case where $u_1=u_2=u_3=u$. In this case, the remainder function reads
\begin{align}\label{R62equal}
R_6^{\rm WL, (2)}(u)&=\int_0^1 dt\Big\{\frac{x_0+y_0+z_0}{t}+\frac{x_1+y_1+z_1}{t-1}+\frac{x_e+y_e+z_e}{t-\frac{1}{1-u}}+\frac{y_{e1}+z_{e1}}{t-\frac{1}{u}}+\frac{y_{e2}}{t-\frac{1-u}{1-2u}}\Big\}\ ,
\end{align}
with
\begin{align}
x_0&=\frac{3}{2}\ln\big(\frac{1-(1-u)t}{1-t}\big)\ln\big(\frac{t(1-(1-u)t)}{1-t}\big)\ln\big(1-(1-u)t^2\big)\ ,\non\\
x_1&=\frac{3}{4}\ln u\big(\pi^2+3\ln^2\big(\frac{1-t}{u}\big)\big)-x_e\ ,\non\\
x_e&=\frac{3}{4}\big(\pi^2+\ln\big(\frac{t(1-(1-u)t)}{1-t}\big)\big(3\ln\big(\frac{t(1-(1-u)t)}{1-t}\big)-2\ln{t}\big)\big)\ln\big(1-(1-u)t^2\big)\ ,\non\\
y_0&=\frac{1}{8}\big(\pi^2+3\big(4\ln u\ln(u t(1-t))+\ln(1-(1-u)t)\big(\ln(1-(1-u)t)-4\ln u+2\ln(1-u)-4\ln (u t)\big)\non\\
&+4\ln(1-t)\ln t-2\ln(1-u t)\ln(u t)-2(\ln(u(1-t))-\ln(1-(1-u)t))\ln(1-u-t+2u t)\big)\non\\
&+6\big(\big({\rm Li_2}\big(\frac{-(1-u)(1-t')}{u}\big)-{\rm Li_2}\big(\frac{-(1-2u)(1-t')}{u}\big)-{\rm Li_2}(u t')+{\rm Li_2}\big(\frac{-(1-2u)(1-(1-u)t')}{u^2}\big)\big)\Big|_{t'=0}^{t}\non\\
&-{\rm Li_2}(u)\big)\big)\ln\big(1-t+u t^2\big)\ ,\non\\
y_1&=-y_{e1}=\frac{3}{4}\ln\big(u t\big)\big(\ln(u t)-\ln(1-(1-u)t)\big)\ln\big(1-t+u t^2\big)\ ,\non\\
y_e&=-y_0-y_1-2y_{e2}\ ,\non\\
y_{e2}&=-\frac{3}{4}\big(\ln(u t)-\ln(1-(1-u)t)\big)\big(\ln(u(1-t))-\ln(1-(1-u)t)\big)\ln\big(1-t+u t^2\big)\ ,\non\\
z_0&=\frac{1}{8}\big(-24\ln(u t)\ln^2(1-u t)+\ln(1-u t)\big(-5\pi^2+6\big(\ln^2(u(1-t))+2\ln^2 u\big)+9\ln^2{t}+6\ln{t}\non\\
&\times(8\ln(u(1-t))-\ln u)+6\ln(1-u)(5\ln u+8\ln{t})-6{\rm Li_2}(u)+24{\rm Li_2}(t)\big)-2\ln(1-2u t)\big(\pi^2\non\\
&+6\ln(u t)\big(2\ln((1-u)t)+\ln(u t)-4\ln(1-u t)\big)-12{\rm Li_2}(u t)\big)+6\big(\ln(1-u-t+2u t)-\ln(1-u)\big)\non\\
&\times\big({\rm Li_2}(u)+\ln u\big(\ln\big(\frac{1-u}{1-t}\big)-3\big(\ln(u t)-\ln(1-(1-u)t)\big)\big)+\big(\ln(1-t)-2\ln{t}\big)\ln\big(\frac{1-t}{1-(1-u)t}\big)\non\\
&+{\rm Li_2}\big(\frac{u t}{-1+t}\big)\big)\big)\ ,\non\\
z_1&=\frac{1}{8}\big(\big(\pi^2+12\ln(u t)(\ln(1-t)+3\ln{t})\big)\big(\ln(1-u t)-\ln(1-u)\big)-\big(\ln(1-u-t+2u t)-\ln u\big)\big(\pi^2\non\\
&+6\ln u\ln{t}+9\ln^2{t}-6{\rm Li_2}(1-t)\big)-3\big(\ln(1-u-t+2u t)-\ln(1-u)\big)\big(-\big(\ln(1-(1-u)t)-\ln u\big)\non\\
&\times\big(3\ln(u t)+\ln\big(\frac{t}{1-(1-u)t}\big)\big)+2{\rm Li_2}\big(\frac{-(1-u)(1-t)}{u}\big)\big)+6\big(\ln(1-(1-u)t)-\ln u\big)\big(-\pi^2+{\rm Li_2}(u t)\non
\end{align}
\begin{align}
&+3\ln^2\big(\frac{1-t}{t}\big)-2\ln\big(\frac{1-t}{t}\big)\ln u+\ln(u t)\ln(1-u t)\big)-\frac{158}{15}\pi^2\ln t+6(5\ln(1-u)-2\ln u)\ln^2 t\big)\ ,\non\\
z_e&=-\frac{1}{8}\big(\big(\ln(1-u-t+2u t)-\ln(1-u)\big)\big(\pi^2+9\ln^2(1-t)-3\big(4\ln^2(u t)+\ln{t}(2\ln(1-t)-\ln{t})\big)\non\\
&+6\big(\ln\big(\frac{u}{1-t}\big)+2\ln\big(\frac{t}{1-t}\big)\big)\ln(1-(1-u)t)+9\ln^2(1-(1-u)t)+12{\rm Li_2}\big(\frac{-u t}{1-t}\big)-6{\rm Li_2}((1-u)t)\big)\non\\
&+\frac{1}{2}\ln u (13\pi^2+12\ln^2(1-u)+13\ln^2 u)-2\ln(1-u)(4\pi^2-3{\rm Li_2}(u))\big)\ ,\non\\
z_{e1}&=\frac{1}{4}\ln(1-2u t)\big(\pi^2+6\ln(u t)\big(2\ln((1-u)t)+\ln(u t)-4\ln(1-u t)\big)-12{\rm Li_2}(u t)\big)\ .
\end{align}
In general kinematics where the three conformal cross ratios can differ, we obtain the following result for $R_6^{\rm WL, (2)}$,
\begin{align}\label{R62general}
&R_6^{\rm WL, (2)}(u_1, u_2, u_3)=\int_0^1 dt\Big\{\frac{I_1+J_1+I_4+K_1+K_2+K_3}{t}+\frac{I_2+J_2+I_5+K_4+K_5}{t-1}+\frac{I_3+J_3+I_6+K_6}{t-\frac{1}{1-u_1}}\non\\
&+\frac{I_7+K_7}{t-\frac{1}{u_1}}+\frac{I_8}{t-\frac{1-u_2}{1-u_1-u_2}}+(u_1\to u_2, u_2\to u_3, u_3\to u_1)+(u_1\to u_3, u_2\to u_1, u_3\to u_2)\Big\}\ ,\hspace{-4em}
\end{align}
where the numerators are given by
\begin{align}
I_1&=I_3=\frac{1}{8}\CG(v_{123};t)\Big\{\pi^2+\big(\ln{t}-\ln{(1-t)}+\ln{(1-(1-u_1)t)}+\ln{u_2}-\ln{u_3}\big)\big(3\ln{t}-5\ln{(1-t)}\non\\
&+3\ln{(1-(1-u_1)t)}+\ln{u_2}-\ln{u_3}\big)\Big\}\ ,\non\\
J_1&=-\frac{1}{8}\CG(v_{132};t)\Big\{\pi^2+\big(3\ln{t}-\ln{(1-t)}-\ln{(1-(1-u_1)t)}-\ln{u_2}+\ln{u_3}\big)\big(\ln{t}-\ln{(1-t)}\non\\
&+\ln{(1-(1-u_1)t)}-\ln{u_2}+\ln{u_3}\big)\Big\}\ ,\non\\
I_2&=\frac{1}{4}\Big\{-\CG(v_{123};t)\big(\pi^2+\big(\ln{t}-\ln{(1-t)}+\ln{(1-(1-u_1)t)}+\ln{u_2}-\ln{u_3}\big)\big(2\ln{t}-3\ln{(1-t)}\non\\
&+2\ln{(1-(1-u_1)t)}+\ln{u_2}-\ln{u_3}\big)\big)+\CG(v_{123};1)\big(\pi^2+\big(3\ln{(1-t)}-2\ln{u_1}-\ln{u_2}+\ln{u_3}\big)\big(\ln{(1-t)}\non\\
&-\ln{u_1}-\ln{u_2}+\ln{u_3}\big)\big)\Big\}\ ,\non\\
J_2&=\frac{1}{4}\Big\{\CG(v_{132};1)\ln{u_1}\big(-\ln{(1-t)}+\ln{u_1}-\ln{u_2}+\ln{u_3}\big)-\CG(v_{132};t)\big(\ln{(1-t)}-\ln{t}\non\\
&-\ln{(1-(1-u_1)t)}+\ln{u_2}-\ln{u_3}\big)\big(\ln{t}-\ln{(1-(1-u_1)t)}\big)\Big\}\ ,\non\\
J_3&=\frac{1}{8}\CG(v_{132};t)\Big\{\pi^2-\ln^2{t}+2\ln{t}\ln{(1-(1-u_1)t)}\non\\
&+\big(\ln{(1-t)}-3\ln{(1-(1-u_1)t)}+\ln{u_2}-\ln{u_3}\big)\big(\ln{(1-t)}-\ln{(1-(1-u_1)t)}+\ln{u_2}-\ln{u_3}\big)\Big\}\ ,\non\\
I_4&=\frac{1}{24}\CG(u_{123};t)\Big\{\pi^2+6\big(\CF(u_1,u_2,u_3;t)-\ln{(1-t+\frac{u_1 t}{1-u_2})}\ln{u_2}+2\ln{t}\big(\ln{(1-t)}-\ln{(1-(1-u_1)t)}\non\\
&+\ln{u_2}\big)+\ln{(1-t)}\big(-\ln{(1-(1-u_1)t)}+\ln{u_1}+\ln{u_3}\big)-\ln{u_2}\ln(1-u_2)-{\rm Li_2}(u_2)\big)\non\\
&+3\big(\ln^2{(1-(1-u_1)t)}-2\ln{(1-u_3 t)}\ln{u_3}-2\big(\ln{(1-(1-u_1)t)}-\ln{u_2}\big)\big(\ln{u_1}+\ln{u_3}\big)\big)\Big\}\ ,\non\\
I_5&=\frac{1}{4}\CG(u_{123};t)\big(\ln{t}-\ln{(1-(1-u_1)t)}+\ln{u_1}\big)\big(\ln{t}+\ln{u_3}\big)\ ,\non\\
I_6&=-\frac{1}{24}\CG(u_{123};t)\Big\{\pi^2+6\big(\CF(u_1,u_2,u_3;t)+\ln{(1-t)}\big(\ln{(1-(1-u_1)t)}-\ln{u_1}+\ln{u_3}\big)+\ln{t}\big(\ln{t}\non\\
&-\ln{(1-(1-u_1)t)}+\ln{u_1}+\ln{u_3}\big)-\ln{u_2}\ln(1-u_2)-{\rm Li_2}(u_2)\big)+3\big(-3\ln^2{(1-(1-u_1)t)}\non
\end{align}
\begin{align}
&-2\big(\ln{(1-t+\frac{u_1 t}{1-u_2})}+\ln{u_1}\big)\ln{u_2}+2\ln{(1-(1-u_1)t)}\big(\ln{u_1}+2\ln{u_2}-2\ln{u_3}\big)\non\\
&+2\big(-\ln{(1-u_3 t)}+\ln{u_1}+\ln{u_2}\big)\ln{u_3}\big)\Big\}\ ,\non\\
I_7&=-\frac{1}{4}\CG(u_{231};t)\big(\ln{t}-\ln{(1-(1-u_2)t)}+\ln{u_2}\big)\big(\ln{t}+\ln{u_1}\big)\ ,\non\\
I_8&=-\frac{1}{4}\CG(u_{123};t)\big(\ln{t}-\ln{(1-(1-u_1)t)}+\ln{u_1}\big)\big(\ln{(1-t)}-\ln{(1-(1-u_1)t)}+\ln{u_2}\big)\ ,\non\\
K_1&=\frac{1}{8}\ln{(1-t+\frac{u_1 t}{1-u_2})}\Big\{\ln{(1-t)}\big(\ln{(1-t)}-2\big(\ln{u_1}-\ln{u_2}+\ln{u_3}\big)\big)+\ln{t}\big(-4\ln{(1-t)}\non\\
&+4\ln{(1-(1-u_1)t)}-6\ln{u_2}\big)+2\big(-{\rm Li_2}(t)+{\rm Li_2}((1-u_1)t)-{\rm Li_2}\big(\frac{-(1-u_1)(1-t)}{u_1}\big)+{\rm Li_2}\big(\frac{u_1-1}{u_1}\big)\non\\
&-\ln(1-t)\ln\big(\frac{1-(1-u_1)t}{u_1}\big)+\ln{u_2}\big(-2\ln{u_1}-\ln{u_3}+\ln(1-u_2)\big)+\ln{(1-(1-u_1)t)}\big(\ln{u_1}+\ln{u_2}\non\\
&+\ln{u_3}\big)+{\rm Li_2}(u_2)\big)
\Big\}\ ,\non\\
K_2&=\frac{1}{24}\Big\{12\ln{(1-u_2 t)}\big(2\ln{t}+\ln{u_2}\big)\ln(1-u_1)-\ln{(1-u_1 t)}\big(5\pi^2-9\ln^2{t}-24{\rm Li_2}(t)\non\\
&-6\big(\big(\ln{(1-t)}+\ln{u_1}\big)^2-2\ln{(1-u_2 t)}\ln{u_1}\big)+12\big(\ln{(1-u_2 t)}-2\ln{u_1}\big)\ln{u_2}+9\big(\ln{u_2}-\ln{u_3}\big)^2\non\\
&-6\ln{t}\big(8\ln{(1-t)}-4\ln{(1-u_2 t)}+3\ln{u_1}+6\ln{u_2}\big)+12\big(\ln{t}+\ln{u_1}\big)\ln{u_3}\non\\
&-6\big(4\ln{t}+2\ln{u_1}+\ln{u_2}\big)\ln(1-u_2)+6{\rm Li_2}(u_2)\big)
\Big\}\ ,\non\\
K_3&=\frac{1}{24}\ln{(1-(u_1+u_2)t)}\Big\{-\pi^2-18\ln^2{t}+12{\rm Li_2}(u_1 t)+12\ln{t}\big(2\ln{(1-u_1 t)}-2\ln{u_1}-\ln(1-u_1)\big)\non\\
&+6\ln{u_1}\big(3\ln{(1-u_1 t)}+\ln{(1-u_2 t)}-\ln{u_2}-\ln(1-u_1)-\ln(1-u_2)\big)
\Big\}+(u_1\to u_2, u_2\to u_1)\ ,\non\\
K_4&=\frac{1}{8}\Big\{\big(\ln{(\frac{u_1}{1-u_2})}-\ln{(1-t+\frac{u_1 t}{1-u_2})}\big)\big(\ln{t}\big(3\ln{t}+2\ln{(1-t)}+2\ln{u_3}\big)+2{\rm Li_2}(t)\big)\non\\
&-\ln(1-t+\frac{u_1 t}{1-u_2})\big(\ln{(1-(1-u_1)t)}\big(\ln{(1-(1-u_1)t)}-2\ln{u_1}-4\ln{t}-2\ln{(1-t)}-2\ln{u_3}\big)\non\\
&+2\big(\ln{u_1}\big(2\ln{t}+\ln{(1-t)}+\ln{u_3}\big)+{\rm Li_2}\big(\frac{-(1-u_1)(1-t)}{u_1}\big)-{\rm Li_2}\big(\frac{u_1-1}{u_1}\big)\non\\
&+\ln(1-t)\ln\big(\frac{1-(1-u_1)t}{u_1}\big)-{\rm Li_2}(1-u_1)\big)\big)
\Big\}\ ,\non\\
K_5&=\frac{1}{24}\Big\{6\big(\big(\ln{(1-(1-u_1)t)}-\ln{u_1}\big)\big(-\pi^2+\big(\ln{t}-\ln{(1-t)}\big)\big(3\ln{t}-3\ln{(1-t)}+2\ln{u_1}\big)\big)\non\\
&+\big({\rm Li_2}(u_1 t)+\ln t\ln(1-u_1 t)\big)\big(\ln{(1-(1-u_2)t)}-\ln{u_2}\big)+\ln{(1-u_1 t)}\ln{u_1}\big(\ln{(1-(1-u_2)t)}-\ln{u_2}\big)\big)\non\\
&+\big(\pi^2+12\big(3\ln{t}+\ln{(1-t)}\big)\big(\ln{t}+\ln{u_1}\big)\big)\big(\ln{(1-u_1 t)}-\ln(1-u_1)\big)-\frac{158}{15}\pi^2\ln t\non\\
&+6(5\ln(1-u_1)-2\ln u_1)\ln^2 t
\Big\}\ ,\non\\
K_6&=\frac{1}{8}\Big\{\ln{(1-t+\frac{u_1 t}{1-u_2})}\big(3\ln^2{t}-\ln^2{(1-t)}+2\ln{(1-t)}\ln{(1-(1-u_1)t)}-3\ln^2{(1-(1-u_1)t)}\non\\
&-2\big(\ln{(1-t)}-\ln{(1-(1-u_1)t)}\big)\ln{u_2}+2\big(\ln{(1-t)}-2\ln{(1-(1-u_1)t)}+\ln{u_1}+\ln{u_2}\big)\big(\ln{t}+\ln{u_3}\big)\non\\
&+2\ln{t}\big(\ln{u_1}+\ln{u_3}\big)+2\big(2{\rm Li_2}(t)-{\rm Li_2}((1-u_1)t)+2\big({\rm Li_2}\big(\frac{-(1-u_1)(1-t)}{u_1}\big)-{\rm Li_2}\big(\frac{u_1-1}{u_1}\big)\non\\
&+\ln(1-t)\ln\big(\frac{1-(1-u_1)t}{u_1}\big)\big)-{\rm Li_2}(1-u_1)-\ln{u_2}\ln(1-u_2)-{\rm Li_2}(u_2)\big)
\big)+\frac{1}{6}\big(10\pi^2\ln{u_2}+\ln{u_1}\non\\
&\times\big(4\ln{u_1}\ln{u_2}-14\ln^2{u_1}-23\pi^2-75\ln^2{u_2}+72\ln{u_2}\ln{u_3}\big)+2\ln(1-u_1)\big(8\pi^2-6\ln{u_2}\ln(1-u_2)\non\\
&+9(\ln{u_2}-\ln{u_3})^2-6\ln{u_1}(\ln{u_2}-\ln{u_3})-6{\rm Li_2}(u_2)\big)\big)
\Big\}\ ,\non
\end{align}
\begin{align}
K_7&=\frac{1}{24}\ln{(1-(u_1+u_2)t)}\Big\{\pi^2+18\ln^2{t}-6\big({\rm Li_2}(u_1 t)+{\rm Li_2}(u_2 t)\big)-12\ln(1-u_1 t)\ln{u_1}-6\ln{t}\non\\
&\times\big(2\ln{(1-u_1 t)}+2\ln{(1-u_2 t)}-3\ln{u_1}-\ln{u_2}-\ln(1-u_1)-\ln(1-u_2)\big)-6\big(\ln{(1-u_2 t)}\non\\
&\times\big(\ln{u_1}+\ln{u_2}\big)+\ln{u_1}\big(-\ln{u_2}-\ln(1-u_1)-\ln(1-u_2)\big)\big)
\Big\}+(u_2\to u_3, u_3\to u_2)\ ,
\end{align}
with (for the definition of the variables $\{u,v\}^{\rm (\pm)}_{ijk}$ see Eq.~(4.3) of ref.~\cite{DelDuca:2009au})
\begin{align}\label{logsqroot}
\CG(\{u,v\}_{ijk};z)&=\ln\big(1-\frac{z}{\{u,v\}^{\rm (+)}_{ijk}}\big)+\ln\big(1-\frac{z}{\{u,v\}^{\rm (-)}_{ijk}}\big)\ ,\\
\CF(u_1,u_2,u_3;z)&=\big(-{\rm Li_2}\big(\frac{-(1-u_1-u_2)(1-z')}{u_1}\big)-{\rm Li_2}(u_3 z')+{\rm Li_2}\big(\frac{-(1-u_1-u_2)(1-(1-u_1)z')}{u_1 u_2}\big)\non\\
&+{\rm Li_2}\big(\frac{-(1-u_1)(1-z')}{u_1}\big)\big)\Big|_{z'=0}^{z}+\ln(1-(1-u_1)z)\ln\big(\frac{(1-u_1)(1-u_2-(1-u_1-u_2)z)}{u_1 u_2}\big)\non\\
&+\ln(1-z)\big(\ln\big(\frac{1-(1-u_1)z}{u_1}\big)-\ln\big(\frac{1-u_2-(1-u_1-u_2)z}{u_1}\big)\big)-\ln z\ln(1-u_3 z)\ .\hspace{-.6em}
\end{align}

A few comments are in order. The representation presented above for the two-loop hexagon Wilson loop remainder function applies when all three conformal cross ratios $a, b, c$ are less than 1. The presence of the cyclic terms in Eq.~(\ref{R62general}) ensures the symmetry of the remainder function under interchange of the three conformal cross ratios. We have checked numerically that our result agrees with that in ref.~\cite{DelDuca:2009au} within numeric errors. In particular, we have reproduced all the results available in literature for the three conformal ratios less than 1. For instance, we obtained from our representation that
\begin{align}
R_6^{\rm WL, (2)}(1/2)&=-1.26609\ ,\non\\
R_6^{\rm WL, (2)}(0.547253, 0.203822, 0.88127)&=-1.66619\ ,
\end{align}
which agree very well with the results in~\cite{Drummond:2008aq,DelDuca:2010zg}. We also reproduced the asymptotic form of the two-loop hexagon Wilson loop remainder function when the conformal cross ratios become small. The variables $\{u,v\}^{\rm (\pm)}_{ijk}$ above involve square roots of certain combination of the conformal cross ratios. In the result of~\cite{DelDuca:2009au}, the dependence of the remainder function on the conformal ratios through these square-root-containing variables is very complicated, while in our representation they show up only in simple logarithms. In the regions where the square roots develop an imaginary part, the two logarithms in Eq.~(\ref{logsqroot}) can always be combined into one that does not have any square root. The representation presented here is valid for all conformal cross ratios less than 1, it can be analytically continuated into the complex plane. We found that for the conformal cross ratios larger than 1, the results given in~\cite{Drummond:2008aq,Anastasiou:2009kna} are correctly reproduced by the real part of the present representation. In the case that all three conformal cross ratios are equal, one can find from Eq.~(\ref{R62equal}) that the remainder function develops branch cuts not only on the real $u$-axis (e.g. from the logarithms in the numerators), but also on the upper/lower complex plane (e.g. from the last dilogarithm in the 3rd row of the expression for $y_0$). For the branch cuts developed from the first two terms in Eq.~(\ref{R62equal}), they can be straightforwardly read from the branch cuts of the numerators; while for those developed from the remaining terms, their determination is complexified by the structure of the denominators, and the location of these branch cuts can even not be straightforwardly seen from the integrand.

To summarize, we presented an alternate, more compact representation for the two-loop hexagon Wilson loop remainder function. It is a good starting point to further explore the possibility of systematically fixing the BDS ansatz and the potential link between the remainder function at weak and strong couplings.

\section{Conclusions}\label{conclusion}
The proposed duality between planar gluon MHV amplitudes and light-like Wilson loops in $N=4$ SYM holds at six-point two-loop level, but both the results for the planar gluon amplitude and for the Wilson loop differ from the BDS ansatz by a finite remainder function, which is a function of the conformal cross ratios constructed from the momenta of external gluons. In ref.~\cite{DelDuca:2009au}, an analytical result for the two-loop hexagon Wilson loop remainder function was presented. However, their result is rather lengthy and the dependence on the conformal cross ratios appears in a complicated way. In this paper we provide an alternate, more compact representation for the two-loop hexagon Wilson loop remainder function, both in the special case that the conformal cross ratios are equal and in general kinematics, where in the former case the remainder function has been evaluated analytically at strong coupling, yielding a fairly simple functional form. Our representation is based on the observation that the BDS ansatz already contains terms that are functions of the conformal cross ratios only, such terms exhibit a simple structure when written in an integral form. We also used the result of~\cite{DelDuca:2009au} and the general properties of multiple polylogarithms throughout our derivation.

\section{Acknowledgments}
I would like to thank Gregory Korchemsky for suggesting this subject, and for various discussions and comments on the manuscript. I would also like to thank David Kosower for valuable discussions and comments on the manuscript. This work is supported by the French Agence Nationale de la
Recherche under grant ANR-06-BLAN-0142.

\end{document}